\documentclass{jps-cp}
\usepackage{dsfont, graphicx}
\usepackage{amsmath, braket, wrapfig}
\usepackage[dvipsnames]{xcolor}
\usepackage{soul}
\newcommand{\nik}[1]{\textcolor{black}{#1}}
\newcommand{\vm}[1]{\textcolor{black}{#1}}

\title{Spin Squeezing as a Probe of Emergent Quantum Orders}%
 
\author{Ilija K.\ \textsc{Nikolov}$^{1}$, Stephen \textsc{Carr}$^{1,2}$, Adrian G.\ \textsc{Del Maestro}$^{3,4}$, Chandrasekhar \textsc{Ramanathan}$^{5}$, and Vesna F.\ \textsc{Mitrovi\'c}$^{1}$}

\inst{$^{1}$Department of Physics, Brown University, Providence, RI 02912, USA \\
$^{2}$Brown Theoretical Physics Center, Brown University, Providence, Rhode Island 02912-1843, USA\\
$^{3}$Department of Physics and Astronomy, University of Tennessee, Knoxville, TN 37996, USA\\
$^{4}$Min H. Kao Department of Electrical Engineering and Computer Science, University of Tennessee, Knoxville, TN 37996, USA\\
$^{5}$Department of Physics and Astronomy, Dartmouth College, Hanover, NH 03755, USA}

\email{vemi@brown.edu}

\recdate{August 4, 2022}

\abst{Nuclear magnetic resonance (NMR) experiments can reveal local properties in materials, but are often limited by the low signal-to-noise ratio.
Spin squeezed states have an improved  resolution  below the Heisenberg limit in one of the spin components, and have been extensively used to improve the sensitivity of atomic clocks, for example \cite{PhysRevLett.83.2274}.
Interacting and entangled spin ensembles with non-linear coupling are a natural candidate for implementing squeezing. Here, we propose measurement of the spin-squeezing parameter that itself can act as a local probe of emergent orders in quantum materials. In particular, we demonstrate how to investigate an anisotropic electric field gradient via its coupling to the nuclear quadrupole moment.  While squeezed spin states are pure, the squeezing parameter can be estimated for both pure and mixed states. We evaluate the range of fields and temperatures for which a thermal-equilibrium state is sufficient to improve the resolution in an NMR experiment  and probe relevant parameters of the quadrupole Hamiltonian, including its anisotropy.}

\kword{Spin squeezing, Quadrupole moment, Electric field gradient, NMR}

\begin{document}
\maketitle
\section{Introduction}
Developing local probes of matter is relevant not only for basic research in quantum materials but also for generating and exploiting specific properties. Nuclear magnetic resonance (NMR) is one of the forefront spectroscopic techniques for microscopic study of  magnetic systems   and investigation of quantum phases of matter \cite{Lundin2007, Ramesh2010, Berthier2017, Karlsson2022}. Nevertheless, because much of the NMR technique relies on a clear resolution of the spectral lines, complex and unresolved lines become a real obstacle. An important example is Ba$_2$NaOsO$_6$ (BNOO), a Mott insulator with strong spin-orbit coupling (SOC), which is believed to host a complex multipolar-ordered phase, that in the intermediate temperatures is characterized by broadening of the NMR spectra \cite{Lu2017, Liu2018, Liu2018a, Cong2019, Willa2019, Cong2020}, preventing a determination of the exact microscopic nature of this exotic phase of matter.  

In high-precision quantum metrology, measurement resolution is increased through squeezed states generated by non-linear operations, and thus they are sensitive to rotations \cite{Kitagawa1993, Wineland1994, Ma2011}.
 Here, we propose an enhanced NMR probe   using squeezing techniques, depicted in \mbox{Fig.\ \ref{fig:schematic}a}. Specifically, we show how spin squeezing can enable probing of the microscopic nature of complex emergent orders, even when no specific features can be resolved in traditional NMR spectroscopic measurements.    Previous work has detailed either the quadrupole coupling in NMR \cite{Man2006}, or the squeezing parameter of nuclei in electric field gradients \cite{AksuKorkmaz2016}, but the two were not put together in the context of probing complex orders by NMR as presented here. We evaluated the performance of our proposed technique using PULSEE \cite{Candoli2021}. 

\section{Squeezing as an enhanced NMR probe}
\vm{We will now examine how much squeezing different initial spin states produce under the quadrupole Hamiltonian. } 
\subsection{Coherent spin states and the squeezing Hamiltonian}

\begin{wrapfigure}[]{r}{0.56\textwidth}
	\centering
			\vspace{-25pt}
	\includegraphics[width=0.55\textwidth]{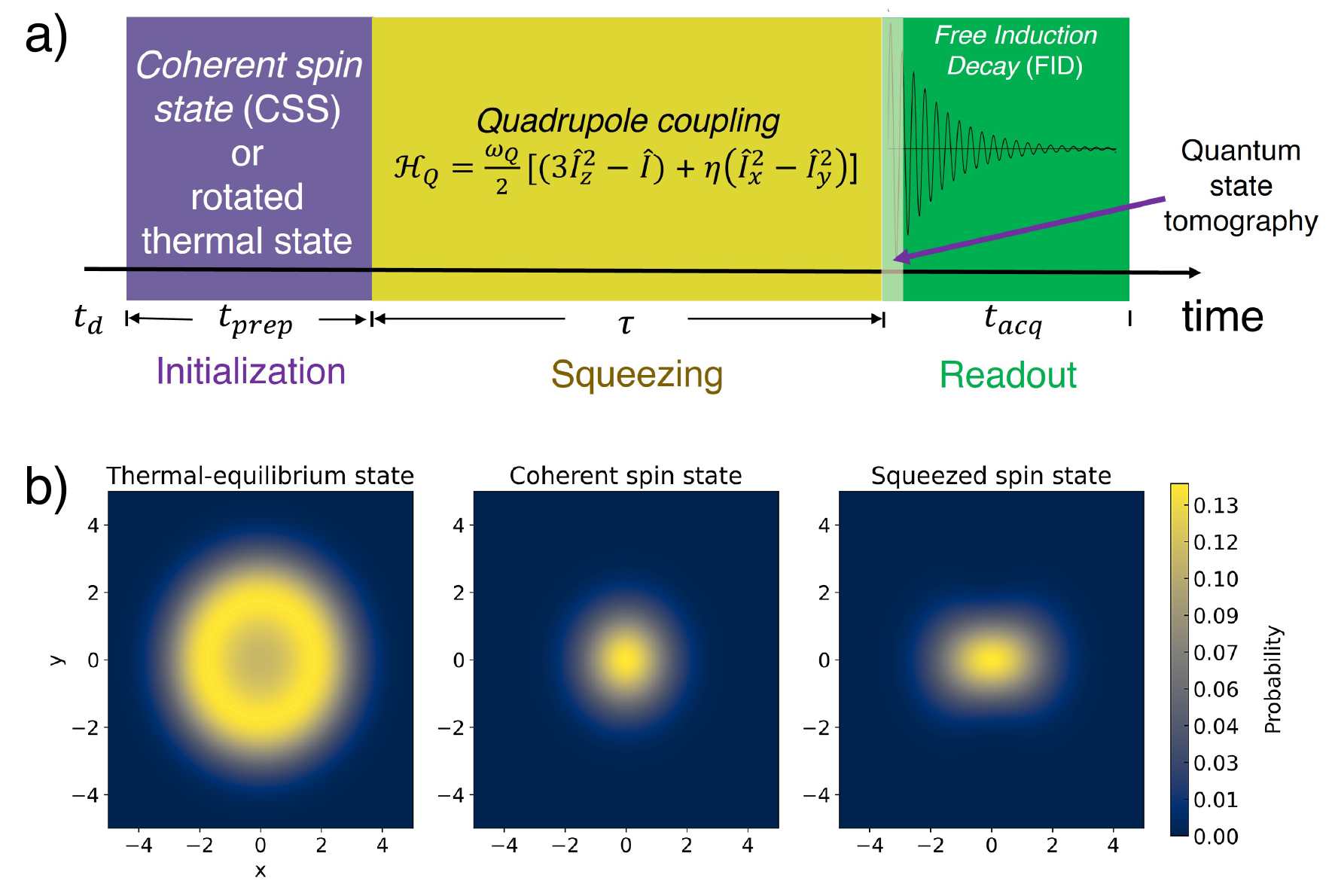}
	\caption{ (\textbf{a}) Schematic of the squeezing NMR probe. While coherent spin states (CSS) give best results, here, we show that a rotated thermal-equilibrium state is sufficient at certain fields and temperatures. (\textbf{b}) Husimi $Q$ function for the initial thermal-equilibrium state $\hat{\rho}_0$ of Eq.\ \ref{eq:thermal}, CSS $\ket{\zeta(0,0)}$, and spin squeezed state (SSS), by the Hamiltonian \mbox{Eq.\ \ref{eq:hamilFullQ}} ($\eta =1 $) at time $t = 0.5\omega_Q^{-1}$. The $Q$ function is obtained by taking $\bra{\alpha} \rho \ket{\alpha}$ for $\alpha = x + iy$.} 
	\label{fig:schematic}
			\vspace{-37pt}
\end{wrapfigure}

 Coherent spin states (CSS) are an eigenstate of the spin momentum operator in a given direction ($\theta, \phi$) that saturate the Heisenberg uncertainty relation \cite{Ma2011}. In terms of the eigenstates of $\hat{I}_z$, CSS are  defined as
$$
{\ket{\zeta(\theta_0, \phi_0)} = \sum^I_{m=-I} \binom{2I}{I+m}^{\frac{1}{2}} \cos(\theta_0/2)^{I+m}\sin(\theta_0/2)^{I-m}e^{i(I-m)\phi_0}\ket{I,m}},$$ where $I$ is the nuclear spin number. The CSS can also be written as $\ket{\alpha} \propto e^{\alpha I_{-}} \ket{I,I}$ for $\alpha = \tan(\tfrac{\theta_0}{2})e^{i\phi_0}$. 
A spin squeezed state (SSS) has a correlated variance that is smaller than the Heisenberg limit in one spin component, at the expense of another non-commuting spin component \cite{Ma2011}. The Husimi $Q$ function is used to illustrate the difference between CSS and SSS in \mbox{Fig.\ \ref{fig:schematic}b}. Any single spin-1/2 system is a one-elementary-spin CSS, and thus cannot be correlated and squeezed \cite{Kitagawa1993, Ma2011}. However, quadrupole nuclei, $I>1/2$, are a natural candidate for producing SSS because a nonlinear spin interaction gives nontrivial quantum correlations between neighboring nuclear spins.

Local symmetry breaking, oftentimes caused by crystalline lattice distortions  and complex multipolar order, may remove rotational symmetry and induce a non-symmetric electronic charge distribution. 
Such a non-zero electric field gradient (EFG) couples  to a nuclear quadruple moment, thus affecting an NMR observable, and becoming a sensitive local probe. 
The quadrupole coupling in the principal axes (PAS) of the EFG is given by
 \begin{equation}
 	\label{eq:hamilFullQ}
 	\mathcal{H_Q}  = \frac{\omega_Q}{2}\Big[(3\hat{I}_z^2-\hat{\mathds{1}})+\eta(\hat{I}^2_x-\hat{I}^2_y)\Big],
 \end{equation}
where $\omega_Q$ is the coupling strength, and the NMR splitting between peaks is $\nu_Q = 3\omega_Q/2\pi$ for $I=3/2$ \& $\eta = 0$.  

\subsection{Thermal-equilibrium states}
In order to extract useful information about the microscopic nature of the material via spin squeezing protocols, the initial state needs be as close as possible to an ideal CSS. However, one often starts an NMR experiment with a mixed thermal-equilibrium state. 
One way to achieve an efficient squeezing from natural thermal-equilibrium states is to use complicated pulse sequences that produce pseudo-pure states of the form $\hat{\mathds{1}} + \epsilon\ket{\psi}\bra{\psi}$. Firstly, we explore how the simplest pulse sequence, in combination with variable temperature and applied field, affects the polarization of the initial state, \textit{i.e.} efficiency of the squeezing.

When working with an ensemble of spins, one can utilize the pseudo-pure state formalism to describe the system using deviation density matrices \cite{Oliveira2011, AuccaiseEstrada2013}. In a Zeeman dominant regime, more spins would be aligned with an applied magnetic field along the $+\hat{z}$ direction, and thus the initial state is
\begin{equation}
    \label{eq:thermal}
	\hat{\rho}_0 = \frac{1}{\mathcal{Z}}\exp(-\frac{\mathcal{H_Q}}{k_BT})\approx \hat{\mathds{1}} - \epsilon \hat{I}_z,
\end{equation}
where $\mathcal{Z}=\Tr\Big(e^\frac{-\mathcal{H_Q}}{k_BT}\Big)$, and $\epsilon$ is the nuclear spin polarization factor which depends strongly on the temperature and magnetic field strength, as well as other characteristics of the material. For a large $\epsilon$, at lower temperatures and higher magnetic fields, if the thermal-equilibrium state is rotated by a $\pi/2$ pulse along $\hat{I}_y$, the deviation density matrix obtained is $\Delta\hat{\rho}\approx-\hat{I}_x$, which we call it a rotated thermal-equilibrium state (RTES), and is our reference for the fidelity of a non-squeezed state. 

Optimal squeezing is achieved with the CSS  $\ket{\zeta(\theta_0 = \pi/2, \phi_0 =\pi)}$, an eigenstate of the $-\hat{I}_x$ operator \cite{Ma2011}. In fact, pseudo-pure nuclear spin coherent states (NSCS) of the form $\hat{\mathds{1}}-\epsilon\Delta\hat{\rho}$, where $\Delta\hat{\rho}=\ket{\zeta(\pi/2, \pi)}\bra{\zeta(\pi/2, \pi)}$ have been previously achieved at room temperature using elaborate methods, such as the strongly modulating pulse technique \cite{Auccaise2015}. Another method to enhance the Zeeman magnetization at room temperature is dynamical nuclear polarization (DNP) \cite{Ramanathan2008}. However, it is a challenge to keep particles of high-temperature systems correlated for a long time. We propose finding an optimal combination of field and temperature for a given material that achieves a nearly-pure, practically maximally ``squeezable'' initial state.

\begin{wrapfigure}[]{r}{0.55\textwidth}
	\begin{center}
		 \vspace{-25pt}
		\includegraphics[width=0.53\textwidth]{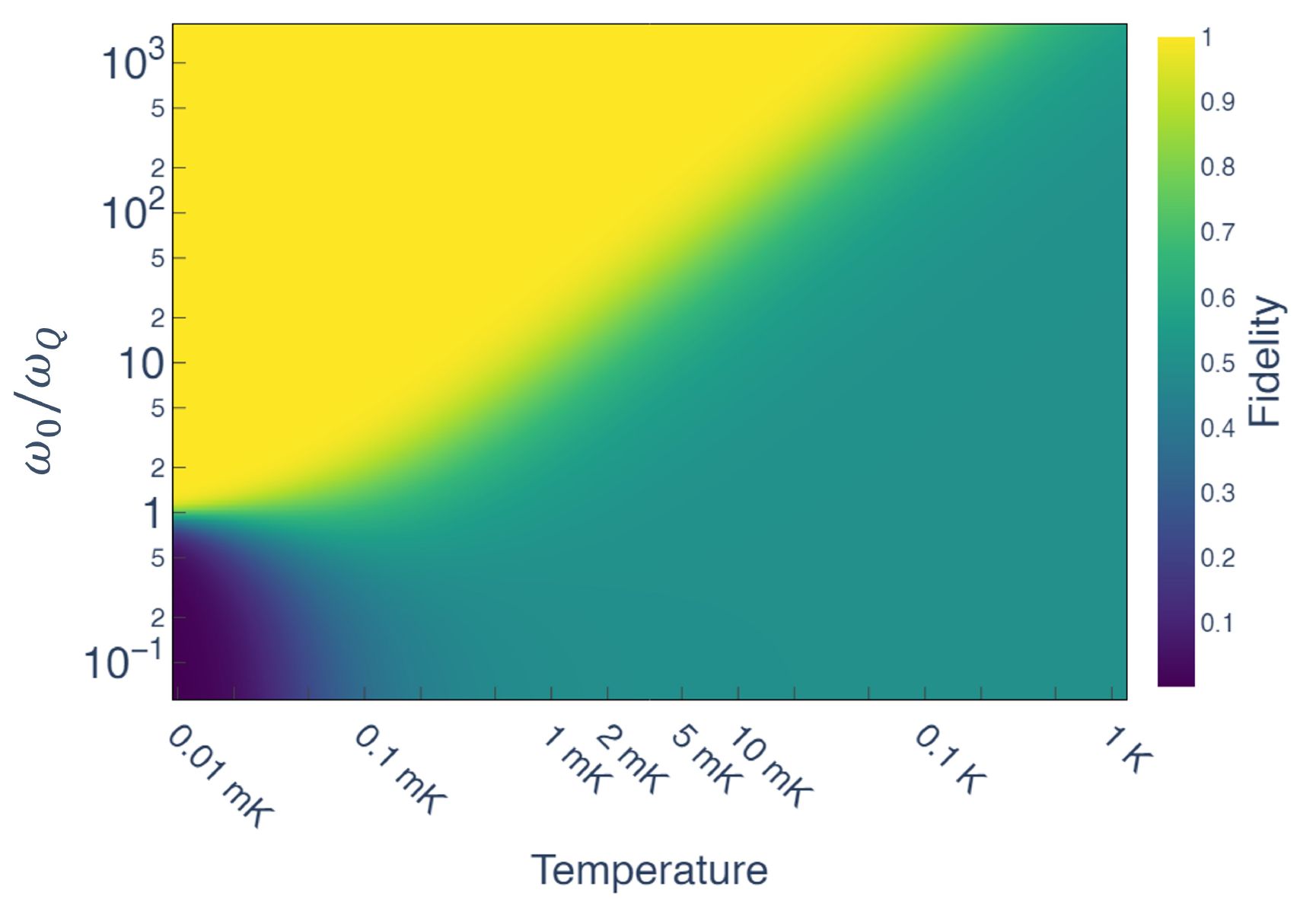}
		\caption{\nik{Illustration of optimal ``squeezability'' of the initial rotated thermal-equilibrium state (RTES) in the parameter space representation given by the} overlap of RTES $\hat{\rho}$ with CSS $\hat{\zeta}$ \nik{(best squeezing)} as a function of magnetic field and temperature. Squeezing is possible for Zeeman-dominant (yellow), and pure NQR (dark blue). } 
		\label{fig:fidelity}
		\vspace{-39pt}
	\end{center}
\end{wrapfigure}

One can determine the overlap between the desired CSS $\hat{\zeta} = \ket{\zeta(\pi/2, \pi)}\bra{\zeta(\pi/2, \pi)}$ and the RTES $\hat{\rho}$ by using a fidelity 
\begin{equation}
	\mathcal{F} = \frac{\Tr(\hat{\zeta}\hat{\rho})}{\sqrt{\Tr(\hat{\zeta}^2)\Tr(\hat{\rho}^2)}}
\end{equation}
that measures how close in ``direction'' the two density matrices are: $\mathcal{F}=1$ implies identical, $\mathcal{F}=0$ implies orthogonal states \cite{Fortunato2002}. The fidelity is given in \mbox{Fig. \ref{fig:fidelity}}, where it is clear that RTES at high-enough field (yellow) produces a NSCS. Roughly, a greater $\epsilon$ gives a stronger NMR signal, or more polarized nuclear spins that can be correlated and squeezed. There is a threshold at $\mathcal{F}\sim.9$, which separates ``squeezable'' from ``non-squeezable'' states (fuzzy green), that increases linearly with the applied field, \vm{as fixed in the case of polarization}. Pure NQR (dark blue) yields nearly pure states of the operator $I_z^2$, which are orthogonal to $\hat{\zeta}$, and \nik{thus} can also be squeezed. However, this requires temperatures below 0.1 mK. Spins in other thermal-equilibrium states (dark cyan, $\mathcal{F}\sim.5$) cannot be squeezed. For simplicity, here we study the case when $\eta = 0$ because the effects of transverse anisotropy on the fidelity of initial thermal-equilibrium states are negligible.

\subsection{Measurement quantities}
We use Kitagawa and Ueda's \cite{Kitagawa1993} squeezing parameter to quantify the ``squeezing'' for both thermal and pseudo-pure states
\begin{equation}
	\label{eq:xiKitagawa}
	\xi^2  = \frac{\text{min}\big(\Delta \hat{I}_\perp\big)}{I}=\frac{C-\sqrt{A^2+B^2}}{I},	
\end{equation}
where $\hat{I}_\perp$ is the vector normal to the mean spin vector (MSV), ${A = \langle \hat{I}_y^2 - \hat{I}_z^2 \rangle}$, ${B = \langle \hat{I}_y\hat{I}_z} + {\hat{I}_z\hat{I}_y \rangle} $ and ${C = \langle \hat{I}_y^2 + \hat{I}_z^2 \rangle}$ for MSV = $\langle \hat{I}_x \rangle$. Numerical calculations show that when the initial states CSS ${\zeta(\ket{\pi/2,\pi})}$ and RTES, in the Zeeman-dominant regime, evolve under $\mathcal{H_Q}$ of \mbox{Eq.\ (\ref{eq:hamilFullQ})}, $\langle \hat{I}_y \rangle = {\langle \hat{I}_z \rangle = 0}$, and thus MSV = $\langle \hat{I}_x \rangle$, for any time. In other words, determining $\xi$ would amount to performing four measurements to obtain $A, B~\&~C$, usually using quantum tomography \cite{Oliveira2011}. A CSS has ${\xi =1}$ and SSS ${\xi< 1}$.

\section{Comparison with NMR quadrupole spectra}

\subsection{Squeezing and relaxation}

\vm{Next, we discuss the ways in which the squeezing parameter may be used to measure relevant terms of the quadrupole Hamiltonian, and compare the utility and efficiency of the squeezing protocols implemented on two different initial states: an ideal CSS and a thermal-equilibrium state.} 

\begin{figure*}[h]
	\centering
		\vspace{-30pt}
	\begin{tabular}{cc}
		\includegraphics[width=0.91\textwidth]{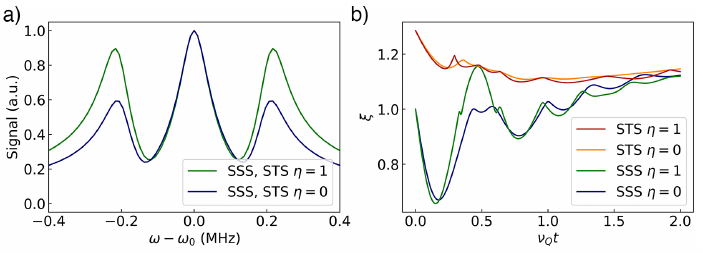}
	\end{tabular}
	\caption{Simulation of relaxation effects. (\textbf{a}) $T_2$-broadened $^{23}$Na spectrum for an aligned sample. Extraction of $\eta$ is experimentally impracticable for both SSS and STS; NMR spectra do not distinguish between these two. (\textbf{b}) The squeezing parameter $\xi$ differentiates different $\eta$ values only for the SSS, and not STS. The quadrupole splitting is $\nu_Q$ = 200 kHz, at 100 mK and 7 T; $\omega_0$ = 78.82 MHz.}
	\label{fig:quadr_sq}
	\vspace{-20pt}
\end{figure*}

In the simulations, we set $T$ = 100 mK, $B_0$ = 7 T. In BNOO, $^{23}$Na is the NMR site of $^{23}\gamma_n$~=~11.26 MHz/T, with a quadrupole splitting $\nu_Q$ = 200 kHz in the low temperature regime \cite{Lu2017}, and the PAS is parallel to the \nik{Zeeman} quantization axis. In order to simulate broadening of the NMR spectra (Fig.\ \ref{fig:quadr_sq}a), we include a phenomenological $T_2$ transverse relaxation in the free induction decay (FID), and a combination of spin-spin and spin-lattice relaxations for the $\xi$ simulations, for which $T_1 = 2 T_2\approx 0.8 \omega_Q^{-1} $. 

The resulting initial RTES has a fidelity $\mathcal{F} \approx.65$, which is not enough to achieve a NSCS \nik{and an effective squeezing}. This is seen through the $\xi>1$ at $t=0$ for the squeezed rotated thermal-equilibrium pseudo-pure states (STS) at both minimal and maximal anisotropy $\eta$, Fig.\ \ref{fig:quadr_sq}b. Since less spin are polarized and available for squeezing, RTES produces effectively no change in the squeezing parameter $\xi$, and thus it cannot easily distinguish between asymmetries in the sample, Fig.\ \ref{fig:quadr_sq}b. In other words, the decay processes suppresses most of the  $\xi$-signal almost instantaneously, at least for this field and temperature. 

The amount of anisotropy shows up mostly in the relative heights in the NMR spectra as the anisotropic correction to the energy is negligible whenever the two relevant \nik{quantization axes} are parallel. However, relative spectral heights in NMR experiments are notoriously difficult to ascertain, and one would usually perform painstaking rotation experiments to determine $\eta$ (Sec.\ \ref{sec:euler}). On the other hand, the two prepared states, STS and SSS, produce identical spectra for the same $\eta$, suggesting that standard experiments cannot provide information about the amount of squeezing, \nik{and hence}, using the squeezing parameter $\xi$, one can readily determine the anisotropy by looking at the reduction of squeezing, ``anti-squeezing,'' at times $t\sim  0.5\nu_Q^{-1}$. When the same relaxation parameters, $T_1, T_2$, are used, CSS detrimentally outperform RTES. Thus, the system must initially be prepared into a NSCS ($\mathcal{F}>.9$).

In general, an EFG starts developing \nik{at the onset of cubic} symmetry breaking, an example of which is the broken local point symmetry (BLPS) ``inter''-phase in BNOO \cite{Lu2017, Liu2018, Liu2018a, Cong2019, Willa2019, Cong2020}. In the BLPS, the strength of the quadrupole interaction is weak, \nik{NMR spectra are not resolved, producing} a broadened central peak \nik{and an expected triplet is not measured}. Thus, little information is obtained about the electric distribution around the nuclear site. When preparing the system in an RTES, our simulations at low temperatures (below \mbox{$\sim 0.1$ K)} suggest that the squeezing parameter $\xi$ can be used to probe the quadrupole splitting and its anisotropy, even with the phenomenological loss of signal. In the particular case of BNOO, one would need to prepare the system in a NSCS as the phase change happens at \mbox{$\sim 10$ K.} 

\subsection{Quadrupole anisotropy with initial CSS}

\begin{figure*}[h]
	\centering
		\vspace{-20pt}
	\includegraphics[width=0.91\textwidth]{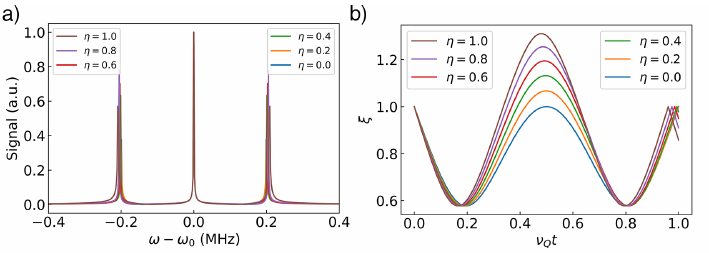} 
	\caption{(\textbf{a}) $^{23}$Na spectra and $\xi$ from $\eta=0$ to $\eta=1$ demonstrating the difficulty in distinguishing between different $\eta$ values for aligned PAS to the applied field. (\textbf{b}) The squeezing parameter $\xi$ shows a strong sensitivity to $\eta$ and has periodicity $\omega_Q$.}
	\label{fig:eta_quad}
	\vspace{-25pt}
\end{figure*}

It is difficult to extract the quadrupole anisotropy parameter $\eta$ from standard NMR spectra, even when preparing a CSS,  $\ket{\zeta(\theta_0, \phi_0)}$, without relaxation and with the PAS aligned to the magnetic field (Fig.\ \ref{fig:eta_quad}a). Nevertheless, the squeezing parameter $\xi$ exhibits sensitivity to changes in the transverse anisotropy $\eta$, both in the low and high $\eta$ limit. As the anisotropy is increased, the squeezing is systematically worsened for $\ket{\zeta(\theta_0, \phi_0)}$, indicated by the increase in the $\xi>1$ region, Fig.\ \ref{fig:eta_quad}b. Namely, one may utilize spin squeezing to extract the quadrupole coupling constant, by looking at the periodicity of the squeezing parameter $\xi$ that depends on $\omega_Q$, and the anisotropy, by $\xi$ detailed measurements around its maximal value at $t\sim0.5\nu_Qt$.

\subsection{Euler angle $\beta_Q$ dependence}
\label{sec:euler}
\begin{figure}[h]
	\centering
	\includegraphics[width=0.97\textwidth]{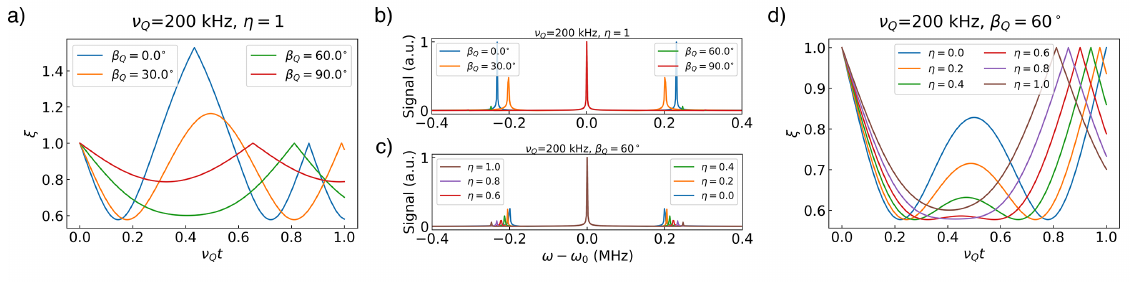}
	\caption{Dependence on the Euler angle $\beta_Q$ up to second order for $\ket{\zeta(\pi/2, \pi)}$. Orientation of the sample can influence the amount of squeezing, and best squeezing (minimal $\xi$) is obtained at the magic angle $\beta_Q \approx 54.74^\circ$ for $\eta=1$. At $\beta_Q=60^\circ$, most squeezing is obtained for $\eta \approx 0.8$. }
	\label{fig:euler}
	\vspace{-20pt}
\end{figure}

As a local probe, NMR can extract information about the orientation of the principles axes of the EFG (PAS) in relation to the LAB frame given by the Euler angles ($\alpha_Q, \beta_Q$ and $\gamma_Q$) up to second order perturbatively. This is given in Fig.\ \ref{fig:euler}a,d where it is clear that $\xi$ is sensitive to both the orientation of the sample $\beta_Q$ and the transverse anisotropy parameter $\eta$. We look at $\alpha_Q = \gamma_Q = 0$, otherwise MSV $\not = \langle \hat{I}_x\rangle$. The central frequency of the satellite peaks in the corresponding NMR spectra is shifted as a function of $\beta_Q$ whenever $\eta =1$. The frequency of the satellite peaks can also change as a function of $\eta$ for angles $\beta_Q\not=$ 0, and Fig.\ \ref{fig:euler}c shows the $\beta_Q = 60^\circ$ case. Whenever $\eta=0$ the central frequency of the satellite peaks does not change as a function of $\beta_Q$, but the squeezing parameter $\xi$ does (not shown).

\section{Concluding remarks}
A spin squeezing method is proposed to locally probe emergent order in systems where standard NMR spectra is unresolved. In particular, by measuring the time evolution of the squeezing parameter $\xi$, one obtains the quadrupole splitting, anisotropy, and the orientation of the \nik{principal axes in the} sample. This proposed method requires the preparation of pseudo-pure nuclear spin coherent states (NSCS). \nik{We show that it may} be done by cooling down the sample and placing it at a high magnetic field, and then applying a standard $\pi/2$ pulse. When one does not have access to either regime, or changing them significantly induces a phase transition, finely-tuned pulse sequences can be used to prepare the system in a NSCS.  Finally, because spin squeezing is a good measure of quantum entanglement~\cite{Ma2011}, our method can be extended to study entanglement in systems with  many-body particle correlation, encompassing both local and collective phenomena.

\section*{Acknowledgments}
We thank William J.\ Kaufman for helpful discussions. This work was supported in part by the US National Science Foundation through Grants OIA-192199 and DMR 1905532. A.D. was supported by the U.S. Department of Energy, Office of Science, Office of Basic Energy Sciences, under Award Number DE-SC0022311.

\end{document}